\documentclass[prd,aps,nofootinbib,twocolumn,showpacs,amsmath,superscriptaddress]{revtex4-2}
\usepackage{graphicx}
\usepackage{comment}
\usepackage{epstopdf}
\usepackage{bm}
\usepackage{epsfig}
\usepackage{graphics}
\usepackage{xspace}
\usepackage{amssymb}
\usepackage{amsmath}
\usepackage{xcolor}
\usepackage[utf8]{inputenc}
\usepackage[normalem]{ulem}
\usepackage{stmaryrd}
\usepackage[colorlinks=true,allcolors=blue]{hyperref}

\def\OMIT#1{}

\newcommand{\nn}{\nonumber}

\newcommand{\bea}{\begin{eqnarray}}
\newcommand{\eea}{\end{eqnarray}}

\newcommand{\beq}{\begin{equation}}
\newcommand{\eeq}{\end{equation}}

\newcommand{\gsim}{\mathrel{\rlap{\lower4pt\hbox{\hskip1pt$\sim$}}\raise1pt\hbox{$>$}}}

\newcommand{\be}{\begin{equation}}
\newcommand{\ee}{\end{equation}}

\newcommand{\APV}[0]{\ensuremath{A_{\rm PV}}}

\begin{document}

\title{\bf  Weak Charge Form Factor Determination at the Electron-Ion Collider}

\author{Hooman Davoudiasl}
\affiliation{
High Energy Theory Group, Physics Department, \\
Brookhaven National Laboratory, Upton, NY 11973, USA
}
\author{Hongkai Liu}
\affiliation{
High Energy Theory Group, Physics Department, \\
Brookhaven National Laboratory, Upton, NY 11973, USA
}
\author{ Sonny Mantry}
\affiliation{Department of Physics and Astronomy, 
University of North Georgia, Dahlonega, GA 30597, USA}               
\author{Ethan T. Neil}
\affiliation{Department of Physics, University of Colorado, Boulder, Colorado 80309, USA}                  

\begin{abstract}
 \noindent
Determining the weak charge form factor, $F_W(Q^2)$, of nuclei over a continuous range of momentum transfers, $0\lesssim Q^2 \lesssim 0.1$ GeV$^2$, is essential for mapping out the distribution of neutrons in nuclei. The neutron density distribution has significant implications for a broad range of  areas, including studies of nuclear structure, neutron stars, and physics beyond the Standard Model. Currently, our knowledge of $F_W(Q^2)$ comes primarily from fixed target experiments that measure the parity-violating asymmetry in coherent elastic electron-ion scattering. Fixed target experiments, such as CREX and PREX-1,2, have provided high-precision weak charge form factor extractions for the $^{48}{\rm Ca}$ and $^{208}{\rm Pb}$ nuclei, respectively.  However, a major limitation of fixed target experiments is that they each provide data only at a single value of $Q^2$. With the proposed Electron-Ion Collider (EIC) on the horizon,  we explore its potential to impact the determination of the weak charge form factor. While it cannot compete with the precision of fixed target experiments, it can provide data over a wide and continuous range of $Q^2$ values, and for a wide variety of nuclei. We show that with data corresponding to an integrated luminosity of ${\cal L} \sim $ 500/$A$ fb$^{-1}$, where $A$ is the nucleus atomic weight, the EIC can significantly impact constraints by lifting degeneracies in theoretical models of the neutron density distribution. Ensuring EIC detector coverage at low $Q^2$ and large negative pseudorapidities will be essential for such $F_W(Q^2)$ measurements.

\end{abstract}

\maketitle

\underline{\textit{Introduction:}} The distribution of protons and neutrons in nuclei, characterized by the electric and weak charge form factors, respectively,  is a long-standing and central question in nuclear physics. It has significant implications for a broad range of areas, including understanding  nuclear structure, atomic physics,  astrophysical phenomena, and physics beyond the Standard Model (SM). For example, precision measurements of nuclear densities for stable nuclei are needed for a precise determination of the nuclear saturation density~\cite{Myers:1966zz,Horowitz:2020evx}, a fundamental parameter corresponding to the density at which nuclear matter is most stable. The saturation density is expected to be universal across stable heavy nuclei due to the short range nature of nuclear forces. Nuclear density measurements can also constrain the form of the isovector component~\cite{Li:2018lpy} of the effective nuclear interaction, sensitive to differences in the distributions of protons and neutrons in nuclei. It is the main driver of nuclear symmetry energy~\cite{Horowitz:2014bja,Baldo:2016jhp} which characterizes the energy required to deform nuclei from a symmetric state of equal numbers of protons and neutrons. Correspondingly, such effects are especially important for understanding the structure and stability of neutron-rich nuclei, shedding light on binding energies, the nuclear drip line, and neutron skin thickness.

On the astrophysical front, neutron stars provide macroscopic realizations of neutron-rich  nuclear matter, where both the SM and General Relativity  are needed to describe the relevant phenomena~\cite{Hagen:2015yea,Fattoyev:2017jql,Horowitz:2019piw,Piekarewicz:2019ahf,Wei:2019mdj,Reed:2021nqk,Li:2021thg}. The isovector nuclear interactions play an important role for understanding properties of neutron stars~\cite{Burgio:2021vgk} which can reach densities that are comparable or greater than the nuclear saturation density in their crust and  several times greater in their core. It affects the equation of state, the symmetry energy, and other properties such as the neutron star stiffness, tidal deformity, and radius. These neutron star properties in turn influence multi-messenger observations~\cite{LIGOScientific:2017ync,Ascenzi:2024wws} in astrophysics and cosmology.

Coherent elastic neutrino-nucleus scattering (CE$\nu$NS)  is proportional to the square of the nuclear weak charge form factor.
CE$\nu$NS~\cite{Freedman:1973yd,COHERENT:2017ipa, COHERENT:2020iec, Abdullah:2022zue, Cadeddu:2023tkp,Danielewicz:2002pu} can probe non-standard neutrino interactions~\cite{Pospelov:2011ha, Harnik:2012ni, Cerdeno:2016sfi, Liao:2017uzy,Dutta:2017nht,Farzan:2018gtr,Denton:2018xmq,Han:2019zkz,AristizabalSierra:2019ykk,Denton:2020hop,Han:2020pff, Shoemaker:2020kji, Herrera:2023xun, Xia:2024ytb, Majumdar:2024dms,DeRomeri:2024iaw,Blanco-Mas:2024ale,Carey:2025byg}. Furthermore, CE$\nu$NS is responsible for the \textit{neutrino fog}~\cite{Billard:2013qya,OHare:2021utq,PandaX:2024muv,XENON:2024ijk},  an irreducible background in direct searches for dark matter that rely on nuclear recoil detection. 
Understanding these various phenomena requires accurate theoretical predictions~\cite{Patton:2012jr,Payne:2019wvy,Yang:2019pbx,Co:2020gwl,Tomalak:2020zfh,Hoferichter:2020osn,VanDessel:2020epd,Piekarewicz:2025lel} of the CE$\nu$NS cross sections, and correspondingly, accurate knowledge of the weak charge form factor.

Given the far-reaching implications of a precise understanding of the distributions of protons and neutrons in nuclei, there has been a significant effort to measure these distributions in a wide range of experiments. 
The distribution of protons is relatively well-known~\cite{Angeli:2013epw} from measurements of  coherent elastic electron-nucleus scattering on a wide range of nuclei. Whereas,
the neutron distributions require weak probes, sensitive to the neutron weak charge.
Currently, the cleanest way to probe neutron distributions in nuclei is through fixed target experiments~\cite{Horowitz:1999fk,CREX:2022kgg,Abrahamyan:2012gp,PREX:2021umo} that measure the parity-violating asymmetry ($A_{\rm PV}$) in coherent elastic electron-ion scattering\footnote{Other probes of neutron distributions can be found in Refs.~\cite{Krasznahorkay:1991zz,Chen:2005ti,Klos:2007is,Brown:2007zzc,Zenihiro:2010zz,Tamii:2011pv,Friedman:2012pa,Tarbert:2013jze,Rossi:2013xha,Hashimoto:2015ema,Zhang:2025voj,Tian:2025wbe,Zhang:2025raf}.}. Such exploration is particularly relevant in light of the proposed Electron-Ion Collider (EIC)~\cite{AbdulKhalek:2021gbh}. This asymmetry is sensitive to the weak charge form factor due to the $Z$-boson exchange contribution which plays the role of the weak probe of neutron distributions. However, a major limitation of fixed target experiments is that they each have access to a few nuclei and provide data on the weak charge nuclear form factor, $F_W(Q^2)$,  only at a single value of momentum transfer, $Q^2$, limiting the breadth of available data\footnote{The G0 experiment has measured parity-violating asymmetries over the
range of momentum transfers $0.12\leq Q^2\leq 1.0$ GeV$^2$, but only for the lightest nuclei, i.e. the proton and deuteron ~\cite{G0:2005chy,G0:2011rpu}.}. For example, the CREX~\cite{CREX:2022kgg} and PREX-1,2~\cite{Abrahamyan:2012gp,PREX:2021umo} experiments have each made high precision extractions of $F_W(Q^2)$ for $^{48}{\rm Ca}$ and $^{208}{\rm Pb}$, respectively, at a single value of $Q^2$. The \textit{ab initio} approach to the neutron distributions of $^{48}{\rm Ca}$ and $^{208}{\rm Pb}$ are presented in~\cite{Hagen:2015yea} and~\cite{Hu:2021trw}, respectively.  
However, determining and testing the distribution of neutrons in the nucleus requires knowledge of the weak charge form factor over the continuous range of momentum transfers, $0 \leq Q^2 \lesssim \mathcal{O}(0.1)$ GeV$^2$.

In this {\it Letter}, we explore the potential of an electron-ion collider environment to provide much needed data on nuclear weak charge form factors over a wide and continuous range of $Q^2$ values, and for a wide variety of nuclei, through measurements of $A_{\rm PV}$ in coherent elastic electron-ion scattering. This is a particularly timely and relevant exploration, given the planned construction of the Electron-Ion Collider (EIC) at the Brookhaven National Laboratory over the next decade. 

The EIC will not be able to directly compete with the precision of fixed target experiments like CREX and PREX-1,2, due to its much lower luminosity. However, its access to a wide and continuous range of $Q^2$ values can allow for weak charge form factor measurements that significantly impact existing constraints from CREX and PREX-1,2. In particular, additional EIC measurements of the weak charge form factor can break existing degeneracies in the parameter space of theoretical models of the neutron density distribution.  We show that the EIC would need to include far-backward detectors and collect 500/$A$ fb$^{-1}$ of data, where $A$ is the nucleus atomic weight, in order to significantly impact constraints derived from the CREX and PREX-1,2 data. 

As long as $s\gg Q^2$, the EIC weak charge form factor measurements will be insensitive to the center of mass energy per nucleon, $\sqrt{s}\sim 20-140$ GeV, up to Coulomb distortion and dispersion corrections~\cite{Horowitz:1998vv,Horowitz:2011qm}. Folding in these corrections, one can combine the requisite data over various runs at the EIC. This also facilitates accumulating data from other planned future electron-ion collider facilities~\cite{Anderle:2021wcy}, operating at different center of mass energies per nucleon, $\sqrt{s}\sim 15-20$ GeV. A second interaction point~\cite{AbdulKhalek:2021gbh,Kim:2025jvr} at the EIC can effectively double the accumulated integrated luminosity. All of these factors can be used toward accumulating the needed integrated luminosity of ${\cal L}\sim$ 500/A fb$^{-1}$. The required electron signal events are predominantly in the far backward direction ($\eta_e \lesssim -4$), motivating the study of additional detector instrumentation in this region.

In addition, the EIC can operate a broad selection of ion beams~\cite{AbdulKhalek:2021gbh}, allowing for measurements of the weak charge form factor for a wide range of nuclei, including those for which  no data exists. Correspondingly, in the Supplemental Material (SuM)~\cite{appendix}, we also provide EIC projection analyses for the extraction of the weak charge form factors for the $^{132}{\rm Xe}$ and $^{40}{\rm Ar}$ nuclei, relevant for dark matter direct search and CE$\nu$NS experiments. The EIC also provides a unique opportunity to study the neutron distributions of unstable radioactive isotopes, which are much less accessible in fixed-target experiments.

\underline{\textit{Background and Formalism:}} In the non-relativistic limit\footnote{For the relativistic case, see Refs.~\cite{Miller:2010nz,Freese:2021mzg}.}, the electric ($F_{\rm e}$) and weak ($F_{\rm W}$) charge form factors  are the Fourier transforms of the electric ($\rho_e$) and weak ($\rho_w$) charge densities in the nucleus, respectively, and can be written as 
\bea
F_i(Q^2) &=& \frac{1}{Q_i}\int d^3r \frac{\sin(Q r)}{Qr}\rho_i(r), \>\> i=\{e,W\},
\label{eq:ff_intro}
\eea
for spherically symmetric density distributions, and we use $Q=\sqrt{Q^2}$. The electric and weak nuclear charges are $Q_{e,W}$.
The corresponding form factors are normalized as $F_i(0)=1$.
The  electric and weak nuclear charges for a nucleus with atomic weight $A$ and  $Z$ protons  are $Q_e=Z e$ and $Q_W= Z Q_W^p + (A-Z) Q_W^n$, respectively. Here $Q_W^p=1-4\sin^2\theta_W \simeq 0.05$ and $Q_W^n=-1$ denote the weak charges of the proton and neutron, respectively, and  $\theta_W$ is the weak mixing angle. Since the neutron is electrically neutral, $F_e(Q^2)$ effectively determines the distribution of protons within the nucleus. By contrast, since the proton weak charge is small, and the neutron weak charge is large,  $F_W(Q^2)$  effectively determines the distribution of neutrons within the nucleus.   

The low-$Q^2$ coherent elastic electron-ion scattering cross section is dominated by single photon exchange, being primarily sensitive to the electric charge form factor. Extracting the weak charge form factor requires isolating  weak interaction effects, mediated by the $Z$-boson, which can be achieved through the asymmetry~\cite{Donnelly:1989qs} 
\bea
\APV(Q^2) \equiv \frac{d\sigma_R/dQ^2-d\sigma_L/dQ^2}{d\sigma_R/dQ^2+d\sigma_L/dQ^2},
\label{eq:apv}
\eea
where $d\sigma_{L(R)}/dQ^2$ denotes the differential cross section with left (right)-handed electrons, and their expressions in the low-$Q^2$ limit are given in the SuM~\cite{appendix}. The asymmetry is sensitive to the interference between the photon and $Z$-boson exchange contributions. This leads to the following expression for the asymmetry~\cite{Donnelly:1989qs}
\beq
\APV \simeq -\frac{G_FQ^2}{4\sqrt{2}\pi \alpha}\frac{Q_W F_{W}(Q^2)}{ZF_{e}(Q^2)},
\label{eq:ApvFw}
\eeq
where the Fermi constant, $G_F$= 1.166$\times 10^{-5}$ GeV$^{-2}$, and the dependence of the asymmetry on the weak charge form factor is manifest.  Note that this is a tree-level formula, omitting loop effects such as Coulomb distortion and dispersion corrections \cite{Horowitz:1998vv,Horowitz:2011qm}.  Such corrections can be significant, particularly at the PREX-1 \& -2 experiments, where the Coulomb distortion correction is roughly 30\% \cite{Horowitz:2011qm}.  These loop corrections have non-trivial dependence on both $Q^2$ and $\sqrt{s}$, and we do not attempt to estimate them for the EIC in this exploratory study.

Characterizing the impact of the form factor extractions on our understanding of the electric and weak charge density distributions through Eq.~(\ref{eq:ff_intro}), requires model parameterizations of $\rho_e$ and $\rho_W$, respectively. While many different parameterizations~\cite{Klein:1999qj,Helm:1956zz,DeVries:1987atn,Co:2020gwl} could be used, we choose the  symmetrized two-parameter Fermi (S2pF) function~\cite{Piekarewicz:2016vbn}: 
\bea
\rho_i(r, c_i,a_i) =  \frac{3Q_i/(4\pi)}{c_i (c_i^2 + \pi^2 a_i^2)}\frac{\sinh{c_i/a_i}}{\cosh{r/a_i}+\cosh{c_i/a_i}}, 
\label{eq:rhoWmodel}
\eea
for $i=\{e,W\}$. The parameters  $c_i$ and $a_i$ correspond to the half-density radius and the surface diffuseness, respectively. 
One of the advantages of this parametrization is that the nuclear form factors can be solved for analytically~\cite{Sprung_1997}
\beq
F_i(Q^2,a,c) = \frac{3\tilde{a}}{\tilde{c}(\tilde{c}^2+\tilde{a}^2)\sinh\tilde{a}}\left[\>\frac{\tilde{a}\sin{\tilde{c}}}{\tanh{\tilde{a}}}-\tilde{c}\cos\tilde{c}\>\right],
\label{eq:ff_S2pF}
\eeq
where we define the dimensionless parameters, $\tilde{a}\equiv\pi Qa$, and $\tilde{c}\equiv Qc$. The impact of potential EIC measurements of $\APV$ will be characterized by the resulting constraints in the $(a_W, c_W)$ parameter space. Similar analyses could be done using different model parameterizations for the electric and weak charge density distributions. Since our main focus is on characterizing the impact of EIC measurements, this simple model parameterization is sufficient. The fiducial $a_W$ and $c_W$ values given in Table~\ref{tab:S2pF} are chosen based on relativistic mean-field theory predictions~\cite{Todd-Rutel:2005yzo,Reed:2020fdf}.

\underline{\textit{Current Experimental Status:}} While the asymmetry in Eq.~(\ref{eq:ApvFw}) is sensitive to the weak charge form factor, it is numerically small due to suppression by the factor of $G_F Q^2$, making it a highly non-trivial and challenging measurement. Correspondingly, very limited data exists. For the $^{48}{\rm Ca}$ nucleus, there is a single precise measurement of $\APV$ from the CREX fixed target experiment at $Q^2 = 0.030\pm 0.0002$ GeV$^2$~\cite{CREX:2022kgg}, yielding $A_{\rm PV}^{\rm CREX} = 2.668 \pm0.113$ ppm and correspondingly, $F_W=0.1304\pm0.0056$. 
For $^{208}{\rm Pb}$, there are two precision measurements from PREX-1~\cite{Abrahamyan:2012gp} and PREX-2~\cite{PREX:2021umo} fixed target experiments.
The PREX-1 and -2 measurements were performed at slightly different momentum transfers, $Q^2 = 0.0088\pm 0.00011$ GeV$^2$ and $Q^2 = 0.0062\pm 0.00005$ GeV$^2$, respectively. The corresponding weak form factor values extracted are, $F_W= 0.204 \pm 0.028$~\cite{Horowitz:2012tj} and $F_W= 0.368 \pm 0.013$~\cite{PREX:2021umo}, respectively. In the analysis to extract these form factor values from the asymmetry data, additional corrections arising from Coulomb distortion effects~\cite{Horowitz:2011qm}, not shown in Eq.~(\ref{eq:ApvFw}), were also included. 

\begin{table}
  \begin{center}
    \begin{tabular}{|c|c|c|c|c|c|} \hline\hline 
     Nucleus & $a^*_{e}$ [fm] & $c^*_{e}$ [fm]  & $a^*_{W}$ [fm] & $c^*_{W}$ [fm] & $Q_{W}$
     \\ \hline                         
         $^{48}{\rm Ca}$ & 0.525 & 3.715 & 0.515 & 3.995 & -26.0 \\  
         \hline
         $^{208}{\rm Pb}$ & 0.512 & 6.666 & 0.614 & 6.815 & -117.9 \\
         \hline  \hline  
    \end{tabular}
        \caption{The fiducial values of the S2pF model parameters ($a_{e},c_{e}$)~\cite{DeVries:1987atn,ANGELI201369} and ($a_{W},c_{W}$)~\cite{Reed:2020fdf},  for $^{48}{\rm Ca}$ and $^{208}{\rm Pb}$, used for making theoretical predictions.}
 \label{tab:S2pF}
  \end{center}
\end{table}

\underline{\textit{Potential Impact of the EIC:}}  Given the relatively limited data available from fixed target experiments, the EIC provides a unique opportunity to provide the critically needed additional data. In the coherent electron-ion scattering process, the recoil electron energy $E_{e'}$ depends on the momentum transfer $Q^2$ and the incoming electron (ion) beam energies $E_e\,(E_A)$. The laboratory frame inelasticity $y\equiv (E_{e'}-E_e)/E_e$ and pseudorapidity $\eta_e$ of the recoil electron can be estimated in the limit of $Q^2\ll E_e^2\ll E_A^2$, we get $y\simeq 0.25 ~Q^2/E_e^2 $ and $ \eta_e \simeq 0.5~\text{log}(y)$.
The currently planned EIC low-$Q^2$ taggers can provide excellent $Q^2$ resolutions for $Q^2 > 10^{-3}$ GeV$^2$. However, the acceptance for $y$ close to zero is very low~\cite{AbdulKhalek:2021gbh,Klest:2025fwx}. Measurements of $\APV$ over the continuous range  $10^{-3}<Q^2 < 0.1$ GeV$^2$ therefore require upgrades in the far-backward detectors. For our analysis, we assume these upgrades have been implemented and set the acceptance to be 100\%. We set the $Q^2$-bin resolution to 5\% for $10^{-3}~\text{GeV}^2 < Q^2 < 10^{-2}~\text{GeV}^2$ and 2\% for $Q^2 > 10^{-2}~\text{GeV}^2$.
 We do not go beyond $Q^2 = 0.1$ GeV$^2$ since there is a physical gap at the EIC in the range $Q^2\sim 0.1$-1 GeV$^2$~\cite{AbdulKhalek:2021gbh}. 

\begin{figure}
	\centering
    {\includegraphics[width=1.\linewidth]{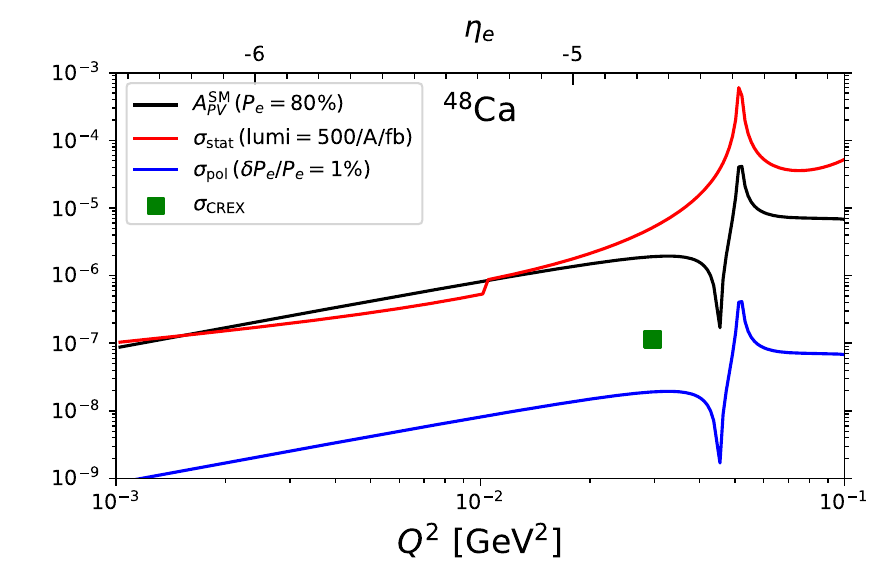}}
    {\includegraphics[width=1.\linewidth]{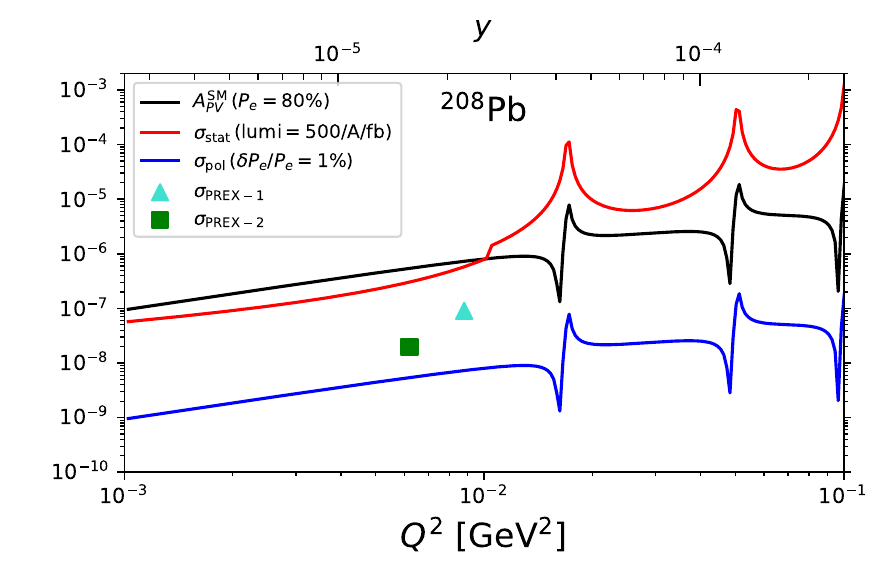}}
	\caption{The SM prediction for $A_{\rm PV}$ (black), together with the expected statistical (red) and systematic (blue) uncertainties from EIC measurements, shown over the $Q^2$ range from 0.001~GeV$^2$ to $0.1$~GeV$^2$. The corresponding values of pseudorapidity $\eta_e$ or inelasticity $y$ are displayed on the top axis for $E_e = 10$~GeV. The uncertainties from the current PREX-1 and -2 measurements are shown as a cyan triangle and green square, respectively. 
	}	
	\label{fig:EIC-FF-band}
\end{figure}

We demonstrate the potential of the EIC for such measurements by considering electron-nucleus coherent scattering with electron and ion beam energies of $E_e = 10$~GeV and $E_A = 110\times A$~\text{GeV}, respectively. The $A_{\rm PV}^i$ value, in the $i$-th $Q^2$ bin, will be diluted by the electron polarization, $P_e$, factor 
 \bea
 A^i_{\rm PV} = P_e \frac{ N^i_{R}-N^i_{L}}{N^i_{R}+N^i_{L}},
 \label{eq:APV_proj}
 \eea
where $N^i_{L(R)}$ are the predicted number of events with a purely left (right)-handed electron beam, and we set $P_e=80\%$. The projected statistical uncertainty in the asymmetry for the $i$-th bin is computed as $\delta A_{\rm PV}^{i, \rm stat.} = \sigma^i_{\rm stat} = 1/\sqrt{N^i_{R}+N^i_{L}}$. We also include a 1\% systematic uncertainty for the asymmetry from the polarization.

Fig.~\ref{fig:EIC-FF-band} shows the theoretical value for $A_{\rm PV}$ (black curve) as a function of $Q^2$ in the S2pF model of the form factors, using the parameters in Table~\ref{tab:S2pF}. These parameters are chosen so that the S2pF model in Eq.~(\ref{eq:ff_S2pF}) reproduces the measured PREX-1,2 and CREX form factor values, and will serve as the fiducial parameters for our analysis. The projected statistical uncertainty for an integrated luminosity of 500/A fb$^{-1}$ is shown by the red curve. A one percent electron beam polarization systematic uncertainty ($\delta P_e= 0.01 P_e, \sigma_{\rm pol}=0.01 \>A_{\rm PV}$) is shown by the blue curve.
We also show the size of the uncertainties on the measurements of $A_{\rm PV}$ in the fixed target experiments CREX, and PREX-1,2. We see that the EIC cannot compete with the precision of the fixed target experiments. It can, however, provide data over a wide and continuous range of previously unmeasured $Q^2$ values.

\begin{figure}
	\centering
    {\includegraphics[width=0.9\linewidth]{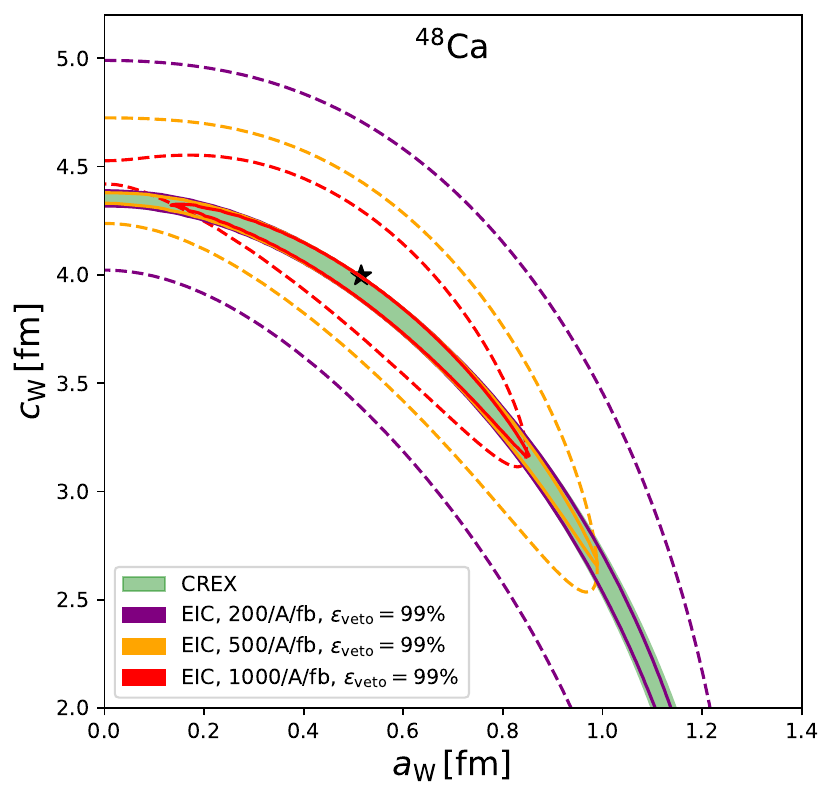}}
    {\includegraphics[width=0.9\linewidth]{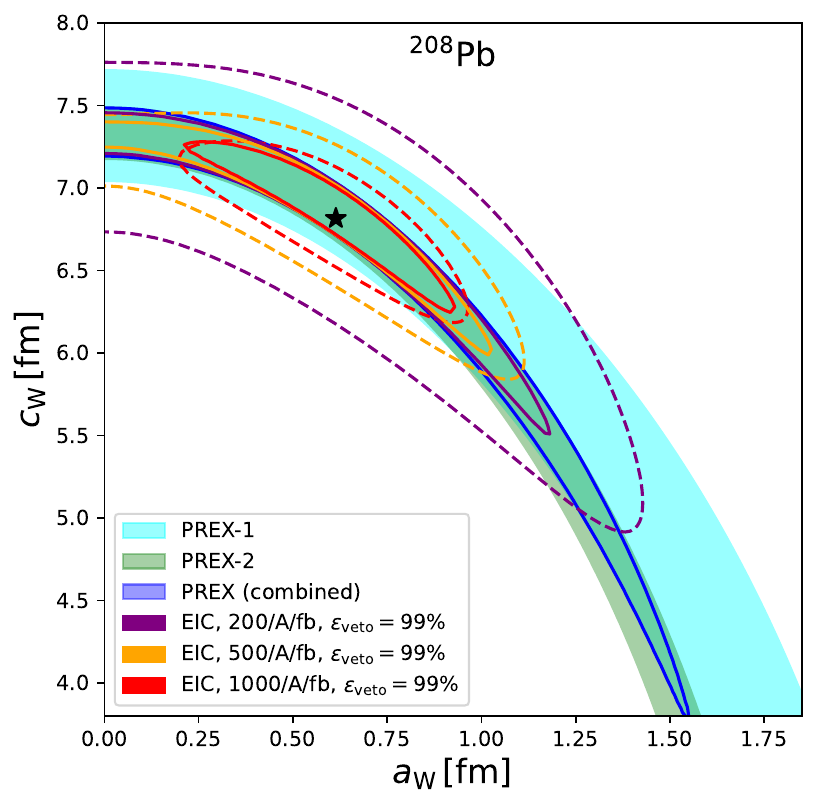}}
	\caption{The 1$\sigma$ limits on $a_{\rm W}$ and $c_{\rm W}$ for $^{48}{\rm Ca}$ (upper panel) and $^{208}{\rm Pb}$ (lower panel). We assume  99\% veto efficiency of incoherent events. The CREX results are shown by green in the upper panel.  In the lower panel, the PREX-1 and -2 results are shown by cyan and green, respectively. The combined PREX result is shown by the solid blue curve. The EIC constraints, assuming $(200,500,1000)$/fb/A integrated luminosity, are shown by dashed (purple,orange,red) contours. The combined results of fixed target experiments and the EIC are shown by the solid contours. The black stars indicate the fiducial values of $a_{\rm W}^*$ and $c_{\rm W}^*$ shown in Table.~\ref{tab:S2pF}.
	}	
	\label{fig:ac_99}
\end{figure}

\begin{figure}
	\centering
	{\includegraphics[width=0.9\linewidth]
    {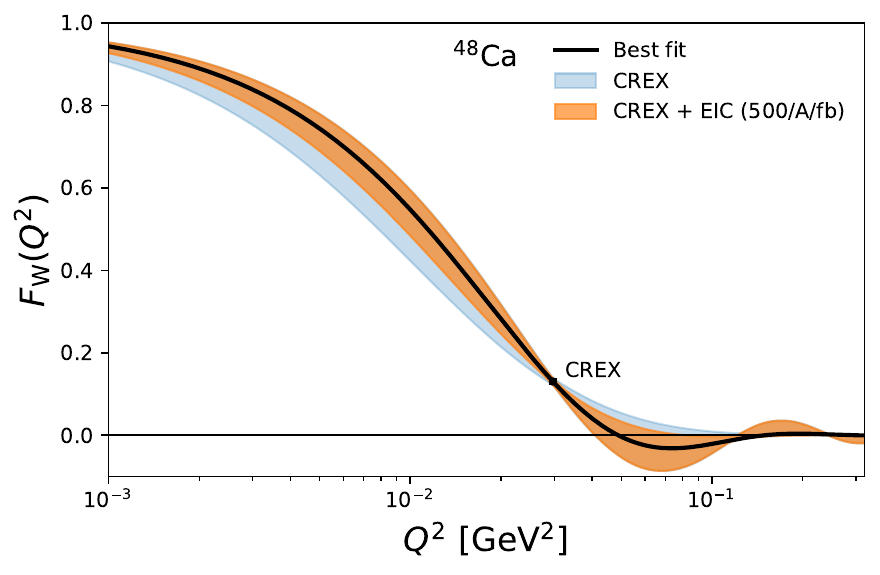}}
    {\includegraphics[width=0.9\linewidth]
    {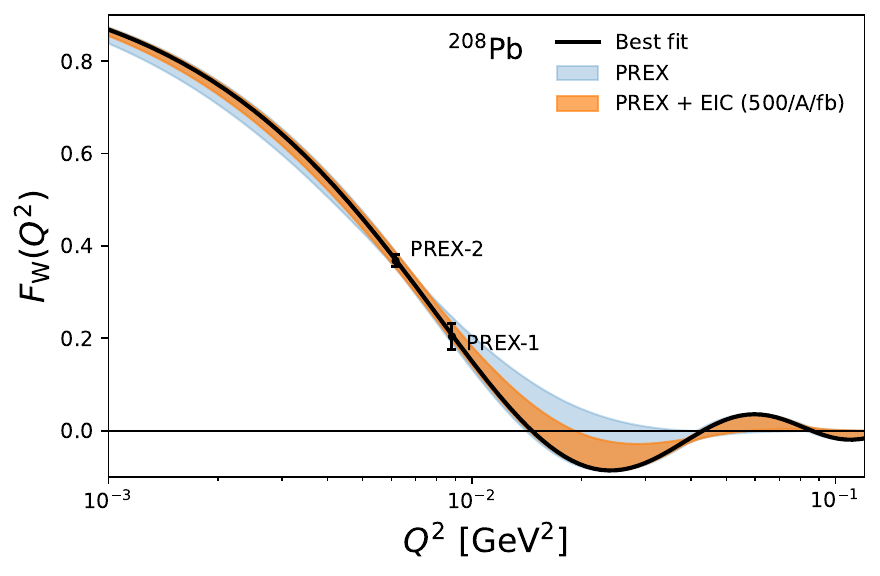}}
	\caption{Constraints on the weak charge form factor in the S2pF model using the 1$\sigma$ allowed region for the $(a_W,c_W)$ parameters shown in Fig.~\ref{fig:ac_99}, using the CREX+EIC and PREX-1+PREX-2+EIC combined data for $^{48}$Ca (upper panel) and $^{208}$Pb (lower panel), respectively.
	}	
	\label{fig:ff}
\end{figure}

From Eqs.~(\ref{eq:ApvFw}) and~(\ref{eq:ff_S2pF}), the theory prediction of $A_{\rm PV}$ effectively depends on the two parameters $a_{W}$ and $c_{W}$, given that $a_{e}$ and $c_{e}$ are well-known. Thus, $\APV$ measurements can be used to constrain the $(a_W,c_W)$ parameter space.  
The $F_W(Q^2)$ extraction from the $\APV$ measurement on $^{48}$Ca at CREX leaves a degeneracy in this parameter space since the measurement is only at a single value of $Q^2$. This is seen in the upper panel of Fig.~\ref{fig:ac_99}, where the 1$\sigma$-contour of allowed values corresponds to the green band. Similarly, the $F_W(Q^2)$ measurements at PREX-1,2  yield the cyan and green 1$\sigma$-contours, respectively, in the bottom panel of Fig.~\ref{fig:ac_99}.  At least two measurements at different $Q^2$ values are necessary to lift the degeneracy in the  $(a_{W},c_W)$ parameter space. However, since the two $Q^2$ values at PREX-1,2 are still quite similar in value, the combined PREX-1,2 results cannot fully lift the degeneracy, as shown by the blue contour.  The EIC can help lift these  degeneracies by providing data at significantly lower and higher $Q^2$ values, as seen in Fig.~\ref{fig:EIC-FF-band}.

Incoherent electron-ion scattering events,  involving ion breakup, are an important source of background at the EIC. It is crucial to tag and veto these  background events with extremely high efficiency, especially at higher $Q^2$.  For our analysis, we assume a veto efficiency of $\epsilon_{\rm veto}=99\%$. Results with $\epsilon_{\rm veto}=99.99\%$, corresponding to the estimated performance of the second interaction point detector~\cite{Aschenauer:2025mku}, show slightly improved but qualitatively similar constraints.

For our EIC analysis, we use the fiducial values for $(a_{W},c_W)$  in Table.~\ref{tab:S2pF}, denoted by the black star in Fig.~\ref{fig:ac_99}. Following the procedure outlined in the SuM~\cite{appendix}, in Fig.~\ref{fig:ac_99}, we show the 1$\sigma$-contour regions, obtained for the integrated luminosities of (200, 500,1000)fb/A at the EIC by the (purple, orange, red) dashed contours, respectively. The corresponding solid contours correspond to 1$\sigma$-contour regions obtained by  combining  the EIC and fixed target experiment data. 
We see  that for integrated luminosities ${\cal L} \sim$ 500/A fb$^{-1}$, the EIC provides non-trivial constraints\footnote{Note that this could be achieved with ${\cal L} \sim$ 250/A fb$^{-1}$ of delivered beam luminosity for two detectors.}, which are further improved when combined with data from fixed target experiments.  Note that the size of the statistical uncertainty scales as $\sigma_{\rm stat}\propto \sqrt{A}/Z$, due to the $Z^2$ scaling of the coherent elastic scattering cross section, dominated by the photon exchange contribution, determining the event counts in each bin, and  the luminosity scaling as ${\cal L}\propto 1/A$. This coherent enhancement for larger nuclei results in more stringent constraints on  $^{208}$Pb compared to $^{48}$Ca, as seen in Fig.~\ref{fig:ac_99}. In Fig.~\ref{fig:ff}, we show the corresponding 1$\sigma$ uncertainty bands for $F_W(Q^2)$, obtained using Eq.~(\ref{eq:ff_S2pF}). We show the uncertainty bands for CREX vs. CREX+EIC data in the top panel and PREX-1+PREX-2 vs. PREX-1+PREX-2+EIC data in the bottom panel. We see that the  EIC  can help significantly improve constraints on the weak charge form factor,  especially for $^{208}$Pb at higher $Q^2$ values, beyond at PREX-1. In the SuM~\cite{appendix}, we show the corresponding results for the $^{132}{\rm Xe}$ and $^{40}{\rm Ar}$ nuclei, relevant for dark matter direct search and CE$\nu$NS experiments.

\underline{\textit{Conclusions:}} A precise determination of the neutron density in nuclei has significant implications for our understanding of nuclear structure, neutron stars, and physics beyond the SM. A clean extraction of the neutron density tends to require weak interaction probes, 
making it experimentally very challenging. Very limited data is available from the fixed target experiments, CREX and PREX-1,2, that have each measured the parity-violating longitudinal electron spin asymmetry in coherent elastic electron-ion scattering to determine the weak charge form factor at a single $Q^2$ value, for $^{48}$Ca and $^{208}$Pb, respectively. Correspondingly, there is a critical need for more data to extract the weak charge form factor of a wide and continuous range of $Q^2$-values, and for a variety of nuclei. The EIC has a unique opportunity to address this critical need. While it cannot compete with the precision
of fixed target experiments, it can provide data over a broad range of $Q^2$ values for various nuclei. For integrated luminosities $\sim$ 500/A fb$^{-1}$, the EIC with upgraded far-backward detectors can provide non-trivial constraints, and lift degeneracies in theoretical models of the neutron density distribution, especially when combined with CREX and PREX-1,2 data.  We emphasize that this measurement requires high electron acceptance in the far-backward direction, which will likely require upgrades and additions to the currently planned detector designs.  This additional detector capability may also be useful for other physics studies, for example the detection of light beyond the Standard Model particles \cite{Davoudiasl:2023pkq,Balkin:2025rtc} and measurement of nucleon axial radius~\cite{Klest:2025bfl}.

\underline{\textit{Acknowledgements:}} We thank Alexander Jentsch and Jiangming Yao for helpful discussions. We also thank Henry Klest for pointing out the importance of detector coverage in the far-backward region. 
The work of H.D. and H.L. is supported by the US Department of Energy under Grant Contract DE-SC0012704.  The work of E.N. is supported by the US Department of Energy under Grant Contract DE-SC0010005.

\bibliographystyle{apsrev4-1}
\bibliography{EIC-elastic-PVES}
\clearpage
\onecolumngrid

\setcounter{equation}{0}
\setcounter{figure}{0}
\setcounter{table}{0}
\setcounter{section}{0}
\renewcommand{\thefigure}{S\arabic{figure}}
\renewcommand{\thetable}{S\arabic{table}}
\renewcommand{\theHfigure}{S\arabic{figure}}
\renewcommand{\theHtable}{S\arabic{table}}
\renewcommand{\theequation}{S\arabic{equation}}

\makeatletter

\begin{center}
	\large{\bf 
		Weak Charge Form Factor Determination at the Electron-Ion Collider
	}\\
	Supplemental Material
\end{center}

\section{Differential cross section}

The  asymmetry, $A_{\rm PV}$, in Eq.~(\ref{eq:apv}), isolates the parity violating contribution from the $Z$-boson mediated weak interaction due to its different couplings to left-handed and right-handed electrons 
\bea
g^e_L &=& \frac{g}{\cos\theta_W}\left (-\frac{1}{2} + \sin^2\theta_W \right), \nn \\
g^e_R &=& \frac{g}{\cos\theta_W}\left ( \sin^2\theta_W\right ), 
\eea
respectively, and $g$ denotes $SU(2)_L$ gauge coupling. In the low-$Q^2$ limit, the differential cross section  is given by 
\beq
\frac{d \sigma_{L,R}}{d Q^2} \simeq \frac{Z^2 e^4}{4\pi Q^4}F^2_{e}(Q^2)\Big [1- \frac{2g^A_Z g^e_{L,R}F_{W}(Q^2)Q^2}{Z e^2F_{e}(Q^2)M_Z^2} \Big ],
\label{eq:sigLR}
\eeq
where the first term is the  photon exchange contribution, proportional to the square of the electric charge form factor.  Note that this is a tree-level formula and does not include loop effects such as Coulomb distortion. The second term inside the square brackets, suppressed by $Q^2/M_Z^2\ll 1$, with $M_Z$ the $Z$ mass, gives the relative correction from the interference of the photon and $Z$-boson exchange contributions. It is sensitive to the weak charge form factor, $F_{W}(Q^2)$, and the vector coupling of the $Z$-boson to the nucleus, 
\beq
g^A_Z = \frac{g}{4\cos\theta_W} Q_W.
\eeq
Using the Eq.~(\ref{eq:sigLR}) for the electron left- and right-handed differential cross sections in Eq.~(\ref{eq:apv}), leads to the expression for the asymmetry in Eq.~(\ref{eq:ApvFw}).

\section{Chi-square analysis}
Here we explain the procedure used to obtain the 1$\sigma$ constraints in Figs.~\ref{fig:ac_99} and \ref{fig:ff}. We derive constraints in the $(a_W,c_W)$ parameter space, as shown in Fig.~\ref{fig:ac_99}, from  measurements of $A_{\rm PV}$ by calculating the chi-square value for each choice of the $(a_W,c_W)$ values as
\bea
\chi^2 &=& \sum_{i=1}^{N_{\rm bin}} \frac{\left [\>A^i_{\rm PV} (a_{W},c_{W})-A^i_{\rm PV}(a^{*}_{W},c^{*}_{W})\>\right ]^2}{(\delta A_{\rm PV}^i)^2},
\label{eq:chisquare}
\eea
where $N_{\rm bin}$ denotes the number of $Q^2$-bins, and $(a_W^*,c_W^*)$ are the fiducial values shown in Table~\ref{tab:S2pF}, used to generate theory predictions for the asymmetry. For $\delta A_{\rm PV}^i$, we include only the statistical uncertainty in each $Q^2$-bin. We ignore the bin-correlated systematic uncertainties arising from the electron beam polarization, since it  is subdominant  to the statistical uncertainty, $\delta A_{\rm PV}^i=\sigma_{\rm stat}^i$, as seen in Fig.~\ref{fig:EIC-FF-band}.  The 1$\sigma$-contour regions are determined by the condition, $\chi^2<2.3 $, as shown in Fig.~\ref{fig:ac_99}.  These 1$\sigma$ regions of allowed $(a_W,c_W)$ parameter values are used in turn to generate the uncertainty bands for 
$F_W(Q^2)$ by using Eq.~(\ref{eq:ff_S2pF}), as shown in Fig.~\ref{fig:ff}.  For analysis of PREX and CREX experimental measurements, we use the experimentally extracted weak form factors $F_W(Q^2)$ as inputs instead.  In this case, the chi-squared function is written in terms of $F_W(Q^2)$ instead of $\APV$ using the tree-level formula tree-level formula Eq.~(\ref{eq:ApvFw}).  As discussed above, this folds in corrections such as Coulomb distortion.

\section{$^{40}$~Ar and $^{132}$~Xe}
\begin{figure}
	\centering
	{\includegraphics[width=0.45\linewidth]{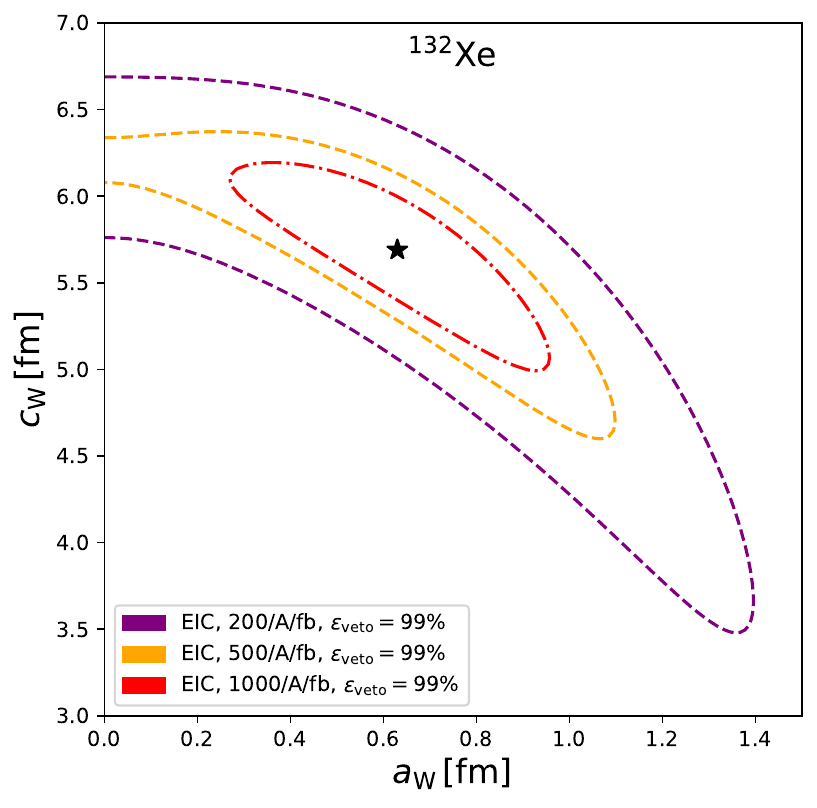}}
    {\includegraphics[width=0.45\linewidth]{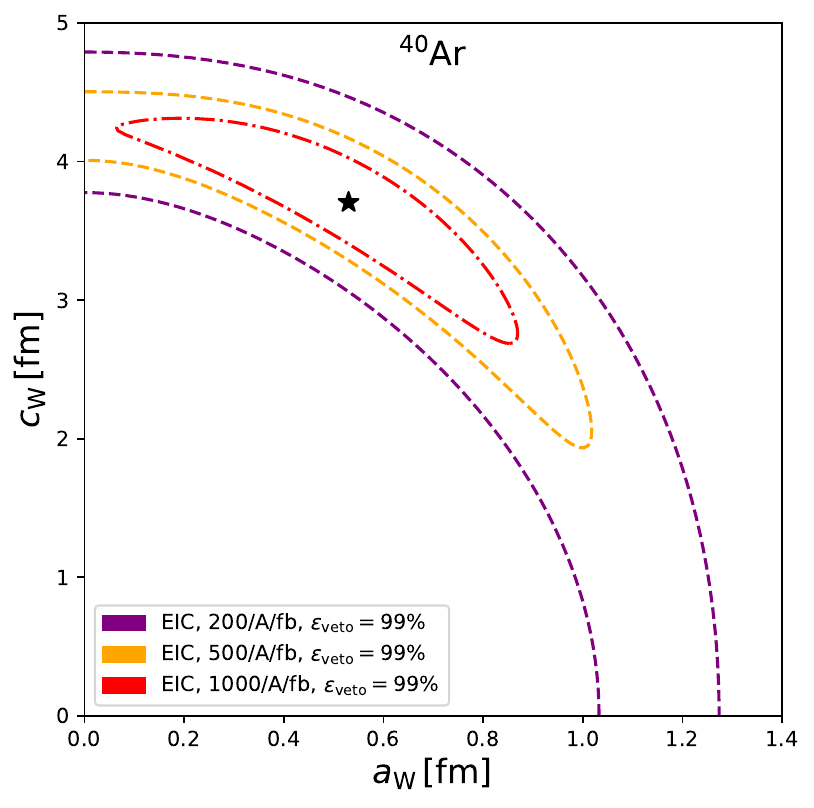}}
	\caption{The 1$\sigma$ limits on $a_{\rm W}$ and $c_{\rm W}$ for $^{132}{\rm Xe}$ (left panel) and $^{40}{\rm Ar}$ (right panel). We assume  99\% veto efficiency of incoherent events.
	}	
	\label{fig:ac_SuM}
\end{figure}

\begin{figure}
	\centering
	{\includegraphics[width=0.45\linewidth]{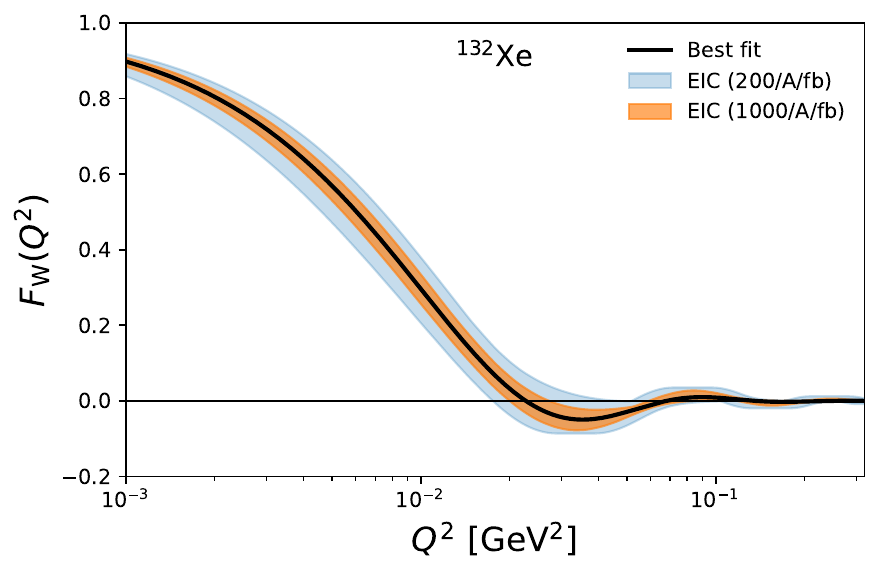}}
    {\includegraphics[width=0.45\linewidth]{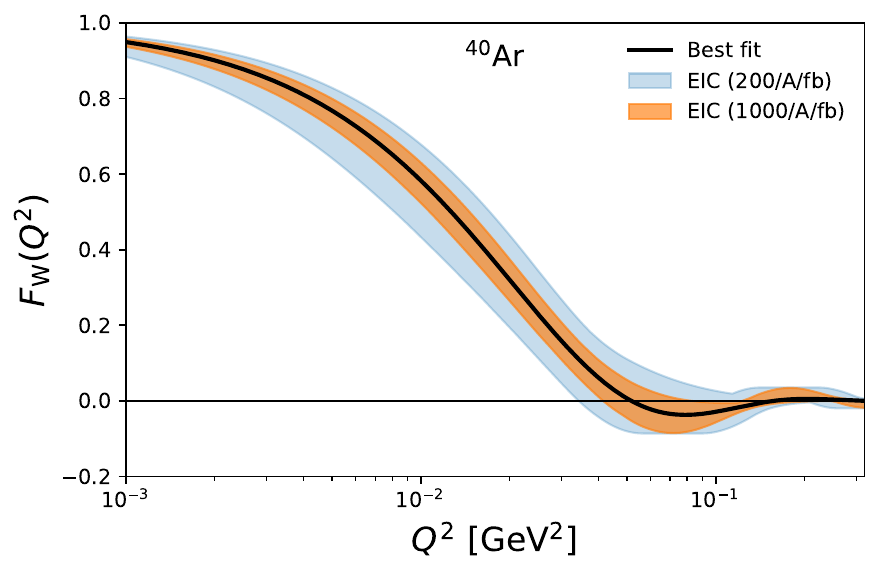}}
	\caption{Constraints  on the weak charge form factor in the S2pF model using the 1$\sigma$ allowed region for the $(a_W,c_W)$ parameters shown in Fig.~\ref{fig:ac_SuM}, using the 200/A/fb and 1000/A/fb data for $^{132}$Xe (left panel) and $^{40}$Ar (right panel), respectively. The veto efficiency of incoherent events is assumed to be 99\%. 
	}	
	\label{fig:ff_SuM}
\end{figure}

In this section, we present additional constraint plots. We present the EIC projections of $^{40}$~Ar and $^{132}$~Xe based on the S2pF parameters given in Table~\ref{tab:S2pF_SuM}. The 1 $\sigma$ limits on $a_W$ and $c_W$ 
are shown in Fig.~\ref{fig:ac_SuM}. The resulting projections on the weak form factors are presented in Fig.~\ref{fig:ff_SuM}. 
\begin{table}
  \begin{center}
    \begin{tabular}{|c|c|c|c|c|c|} \hline\hline 
     Nucleus & $a^*_{e}$ [fm] & $c^*_{e}$ [fm]  & $a^*_{W}$ [fm] & $c^*_{W}$ [fm] & $Q_{W}$
     \\ \hline                         
         $^{40}{\rm Ar}$ & 0.542 & 3.580 & 0.530 & 3.705 & -21.2 \\  
         \hline
         $^{132}{\rm Xe}$ & 0.523 & 5.647 & 0.630 & 5.920 & -75.5 \\
         \hline  \hline  
    \end{tabular}
        \caption{The fiducial values used for the S2pF model parameters ($a_{e},c_{e}$)~\cite{ANGELI201369,Duda:2006uk} and ($a_{W},c_{W}$)~\cite{Co:2020gwl,Piekarewicz:2025lel}  for $^{40}{\rm Ar}$ and $^{132}{\rm Xe}$. }
 \label{tab:S2pF_SuM}
  \end{center}
\end{table}

\clearpage
\onecolumngrid

\end{document}